\documentclass[conference]{IEEEtran}
\IEEEoverridecommandlockouts
\usepackage{cite}
\usepackage{amsmath,amssymb,amsfonts}
\usepackage{algorithmic}
\usepackage{algorithm}
\usepackage{graphicx}
\usepackage{textcomp}
\usepackage{xcolor}
\usepackage{braket}
\usepackage{multirow}
\usepackage{adjustbox}
\usepackage{subfig}

\def\BibTeX{{\rm B\kern-.05em{\sc i\kern-.025em b}\kern-.08em
    T\kern-.1667em\lower.7ex\hbox{E}\kern-.125emX}}
\begin{document}

\title{Residue Number System (RNS) based\\ Distributed Quantum Multiplication\\
}

\author{\IEEEauthorblockN{Bhaskar Gaur} 
\IEEEauthorblockA{\textit{University of Tennessee} \\
Knoxville, TN, USA \\
bgaur@vols.utk.edu}
\and
\IEEEauthorblockN{Himanshu Thapliyal} 
\IEEEauthorblockA{\textit{University of Tennessee} \\
	Knoxville, TN, USA \\
	hthapliyal@utk.edu}
}

\maketitle

\begin{abstract}
Multiplication of quantum states is a frequently used function or subroutine in quantum algorithms and applications, making quantum multipliers an essential component of quantum arithmetic. However, quantum multiplier circuits suffer from high Toffoli depth and T gate usage, which ultimately affects their scalability and applicability on quantum computers. To address these issues, we propose utilizing the Residue Number System (RNS) based distributed quantum multiplication, which executes multiple quantum modulo multiplication circuits across quantum computers or jobs with lower Toffoli depth and T gate usage. Towards this end, we propose a design of Quantum Diminished-1 Modulo (2\textsuperscript{n}+1) Multiplier, an essential component of RNS based distributed quantum multiplication. We provide estimates of quantum resource usage and compare them with those of an existing non-distributed quantum multiplier for 6 to 16 qubit sized output. Our comparative analysis estimates up to 46.018\% lower Toffoli depth, and reduction in T gates of 34.483\% to 86.25\%.
\end{abstract}


\section{Introduction}

Quantum algorithms like Shor’s algorithm, the HHL algorithm, and quantum approximate optimization algorithms depend on quantum arithmetic circuits for representation and optimization \cite{shor1994algorithms, duan2020survey, lykov2023fast}. Quantum multipliers are crucial among these circuits, serving as building blocks for tasks such as factorization, quantum image processing, and quantum cryptanalysis \cite{munoz2018quantum, mohan2016residue, roetteler2017quantum, putranto2022another}. Therefore, optimizing quantum multipliers and making them scalable is a key goal. The Clifford+T gate set is one of the most widely used gate sets, as it can be made fault-tolerant using error-correcting codes \cite{miller2014mapping}. However, because implementing the T gate is expensive, T gate count has become a crucial performance metric for designing fault-tolerant quantum circuits\cite{gosset2013algorithm, gosset2014algorithm}.

Distributed Quantum Computing (DQC) is an emerging computing approach that leverages individual quantum systems to greatly boost computing power and scale quantum algorithms to larger sizes \cite{tang2024distributed, barral2024review}. Residue Number Systems (RNS) is another emerging paradigm that can distribute arithmetic operations across multiple quantum computers or jobs. RNS has demonstrated the ability to perform distributed quantum addition while also making it resilient to crosstalk attacks \cite{gaur2024residue, gaur2024crosstalk}. Since RNS can also achieve closure for multiplication, this work examines the impact of RNS-based distributed quantum multiplication on Toffoli depth and T gate count.
Our approach combines classical computing with quantum computing to distribute multiplication across multiple quantum circuits using Residue Number Systems (RNS). We choose a set of three moduli (2\textsuperscript{n}-1, 2\textsuperscript{n}, and 2\textsuperscript{n}+1), that allows us to create scalable RNS sets with optimized quantum addition implementations that require lower Toffoli depth and T gate count. Although quantum modulo multipliers exist for the moduli 2\textsuperscript{n}-1 and 2\textsuperscript{n}, there was no optimized implementation for the quantum modulo multiplier 2\textsuperscript{n}+1. Hence, we propose a Quantum Diminished-1 Modulo (2\textsuperscript{n}+1) Multiplier (QDMM).

Following are the contributions of our work:
\begin{itemize}
	\item We propose Quantum Diminished-1 Modulo (2\textsuperscript{n} + 1) Multiplier (QDMM), a quantum arithmetic circuit necessary for Residue Number System (RNS) based quantum multiplication.
	\item We provide accurate resource estimates for quantum modulo 2\textsuperscript{n} and (2\textsuperscript{n} - 1) multipliers refined using O(log n) depth estimates based on original QCLA adders by Draper et al \cite{draper2004logarithmic}.
	\item We demonstrate the potential of RNS based distributed quantum multiplication compared to an existing non-distributed quantum multiplier in form of Toffoli depth and T gate count improvements as input size increases.
\end{itemize}


\section{Background}
\label{Background}

\begin{figure}[t]
	\centering
	\includegraphics[scale=0.3]{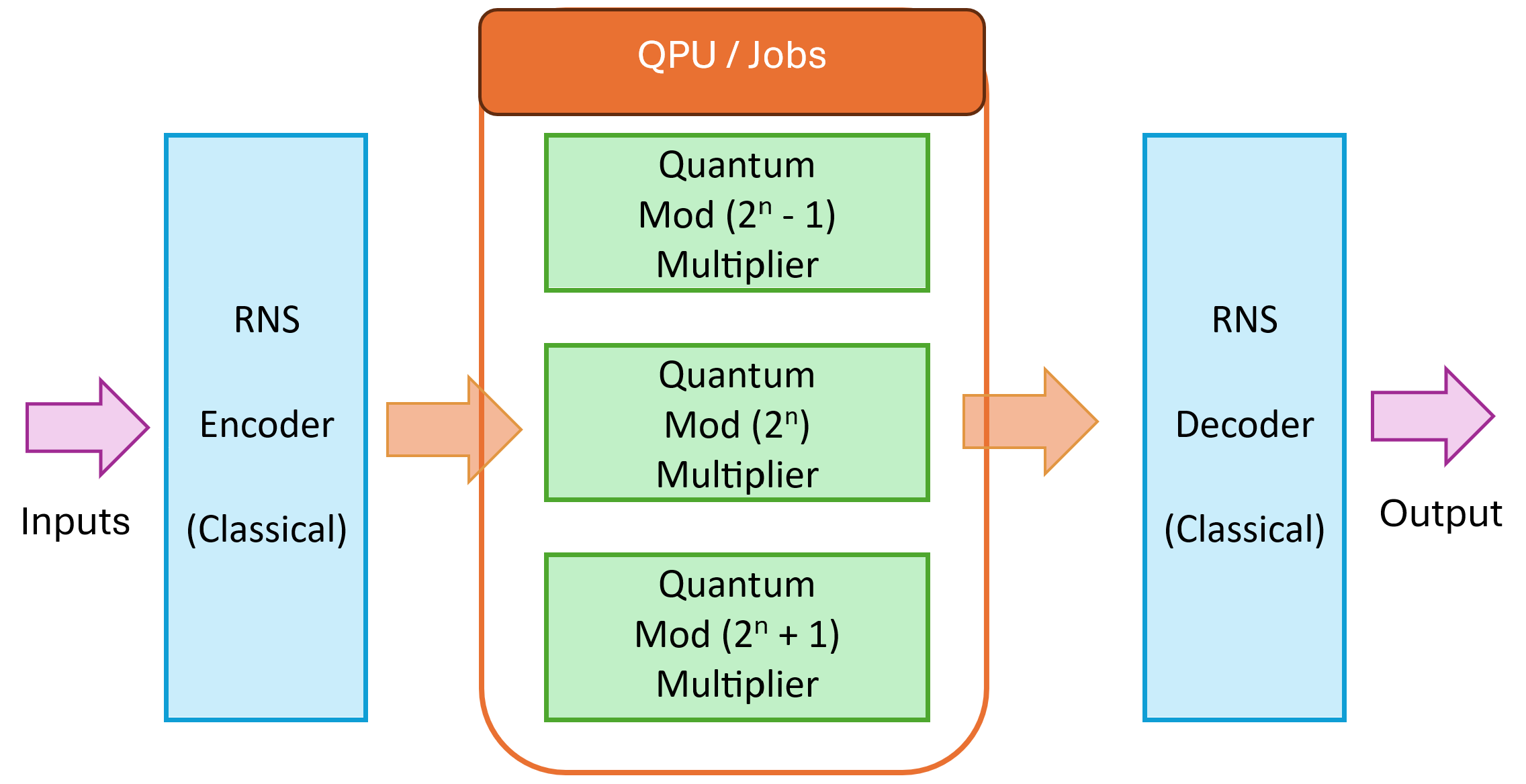}
	\caption{RNS based distributed quantum multiplication based on hybrid classical-quantum flow, that uses parallel independent Residue Number System (RNS) based quantum modulo multipliers distributed across multiple quantum computers/jobs.}
	\label{fig:RNS}
\end{figure}
\begin{figure}[b]
	\centering
	\includegraphics[scale=0.35]{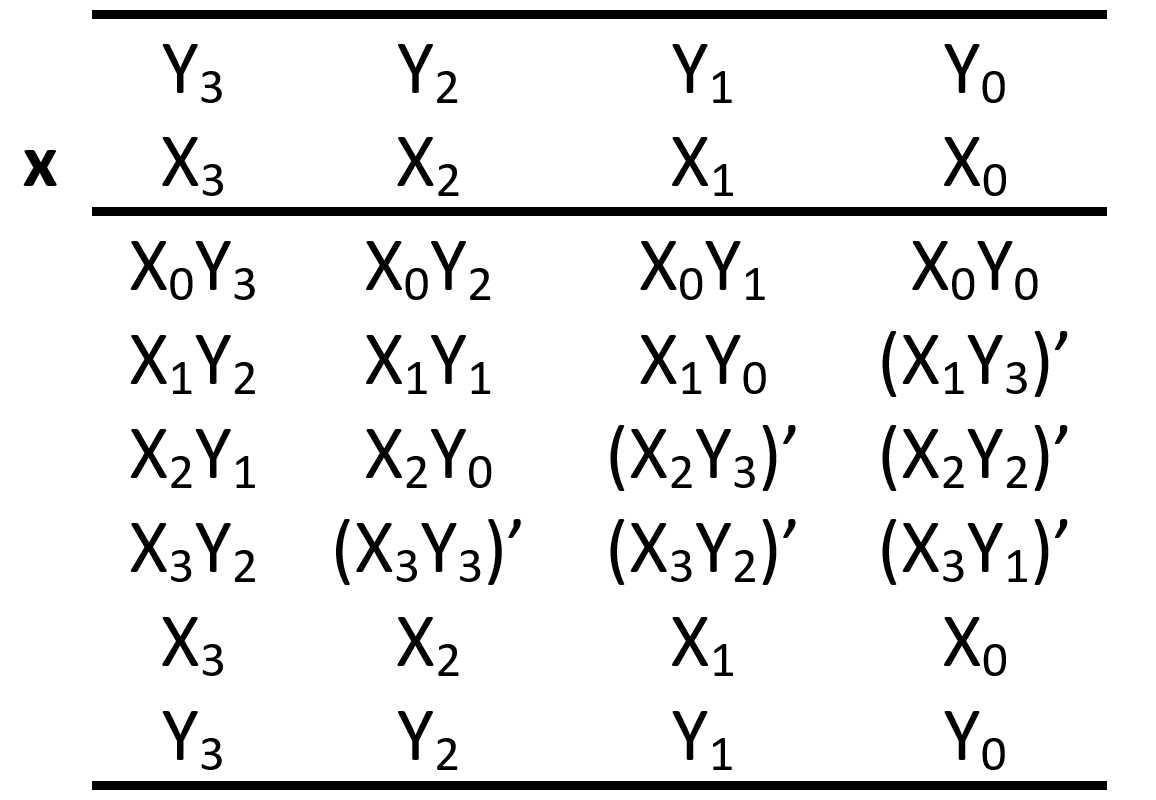}
	\caption{Partial product arrangement for proposed Quantum Diminished-1 Modulo (2\textsuperscript{n}+1) Multiplier (QDMM).}
	\label{fig:mult_pp}
\end{figure}
\begin{figure}[h]
	\centering
	\includegraphics[scale=0.19]{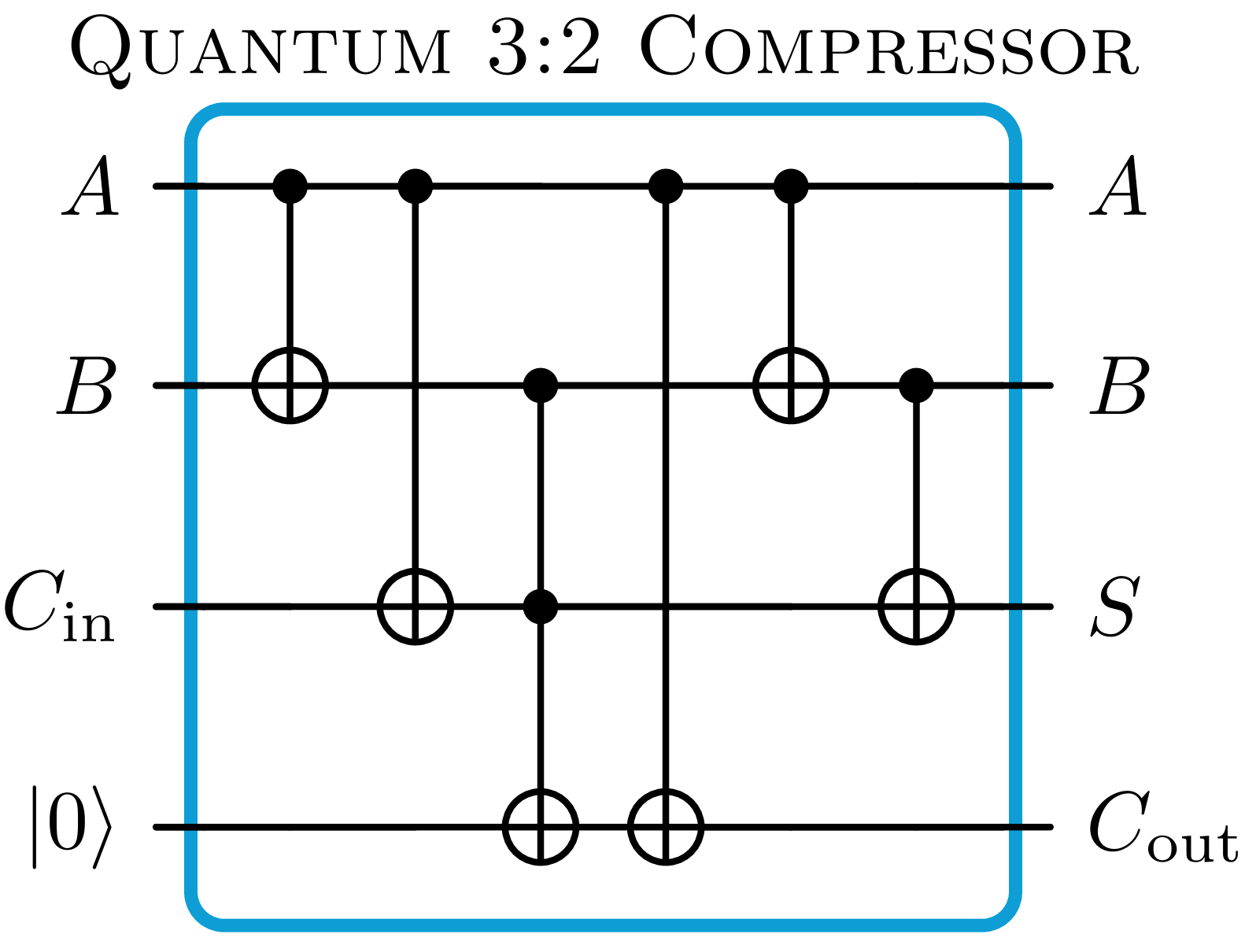}
	\caption{Quantum 3:2 compressor that conducts a single bit full addition for inputs A, B, C\textsubscript{in}, and outputs Sum and C\textsubscript{out}.}
	\setlength{\belowcaptionskip}{1pt}
	\label{fig:qcompressor}
\end{figure}
\begin{figure}[h]
	\centering
	\includegraphics[scale=0.55]{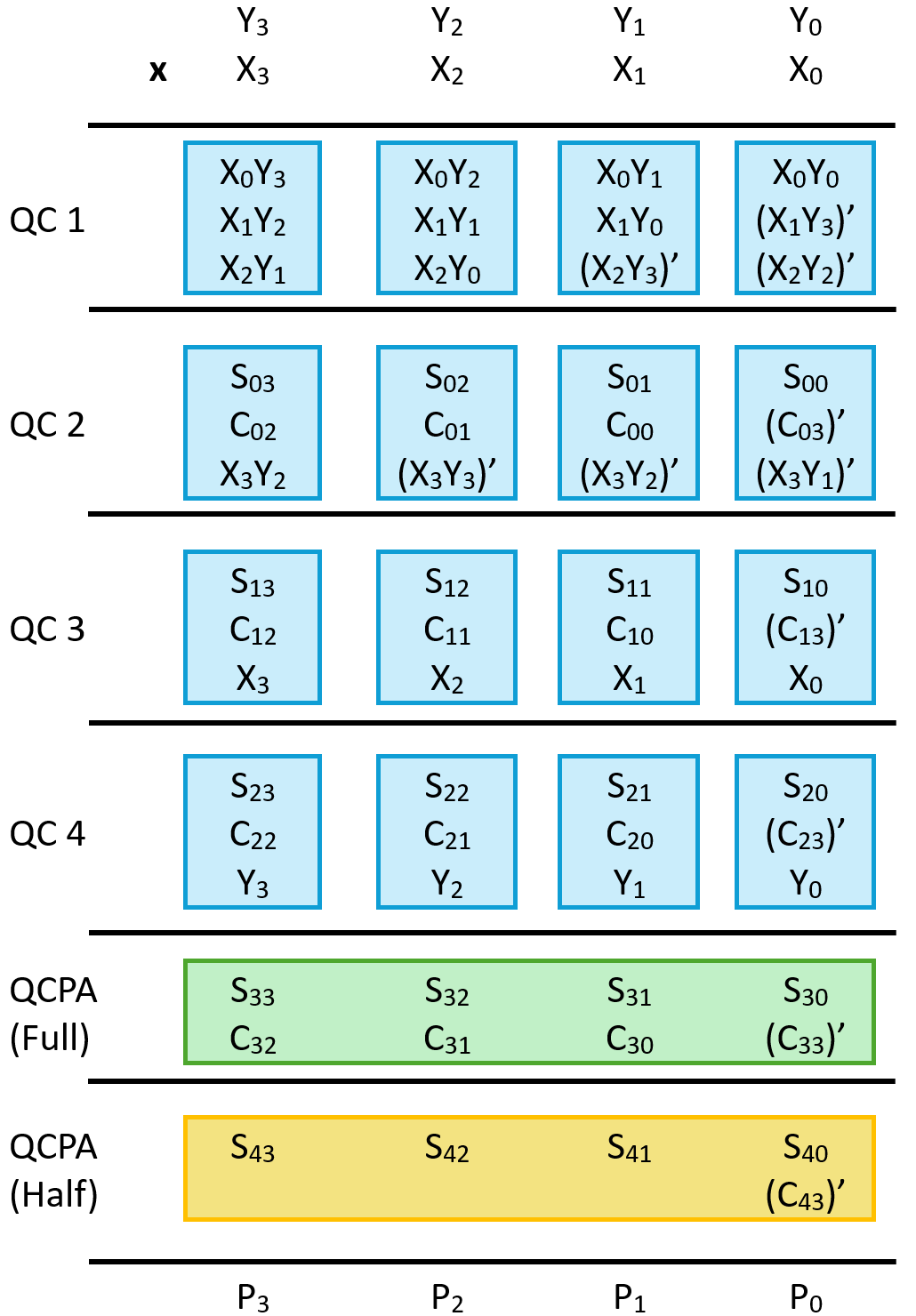}
	\caption{Architecture of proposed Quantum Diminished-1 Modulo (2\textsuperscript{n}+1) Multiplier (QDMM) for (n-1) qubits.}
	\setlength{\belowcaptionskip}{1pt}
	\vspace*{-5mm}
	\label{fig:mult_2n+1}
\end{figure}

\subsection{Residue Number System}

The Residue Number System (RNS) is a numerical representation method in which integers are expressed as their residues modulo a group of relatively prime numbers, known as moduli. This system enables rapid arithmetic calculations, particularly for addition, subtraction, and multiplication, as it processes residues in parallel without relying on carry dependencies \cite{mohan2016residue}. RNS can efficiently scale to larger ranges by choosing moduli that are sufficiently large to cover the intended range while maintaining their relative primality. Typically, moduli in RNS include small prime numbers and powers of two such as (2\textsuperscript{n}-1, 2\textsuperscript{n}, 2\textsuperscript{n}+1). This set allows a more efficient representation of conventional numbers up to nearly 2\textsuperscript{3n}. According to Equation \ref{equation:Range}, the Range of the RNS system is given by the product of the chosen moduli, with M denoting the number of moduli and the set expressed as {R\textsubscript{0}, R\textsubscript{1}, ..., R\textsubscript{M}}. Range is also useful for comparing and selecting among different RNS sets. In this work, the RNS sets provide a range greater than that needed for equivalent multiplication output.
\vspace{-0.2cm}
\begin{equation} \label{equation:Range}
	\begin{aligned}
		&Range = \prod_{1}^{M} R \\
	\end{aligned}
\end{equation}

Figure \ref{fig:RNS} illustrates our method of using a Residue Number System (RNS) in Distributed Quantum Computing (DQC), which converts quantum multiplication into separate quantum modulo multiplier circuits across multiple QPUs or multiple jobs on a single QPU. Unlike other methods such as circuit cutting or quantum teleportation, our approach does not require dependencies between distributed circuits and can be run on contemporary quantum computers \cite{barral2024review, tang2024distributed}.

Classical computing utilizes RNS in various fields, including signal processing, cryptography, and enhancing the reliability of computing systems \cite{mohan2016residue}. In quantum computing, RNS enables larger and faster quantum addition, distributed across multiple machines/jobs, overcoming limitations like qubit count and noise \cite{gaur2024residue}. RNS has shown the potential to secure the quantum addition against crosstalk attacks \cite{gaur2024crosstalk}. RNS has improved resource estimates for quantum factorization and its use in factoring 2048-bit RSA integers \cite{chevignard2024reducing, gidney2025factor}. RNS has the potential to make quantum computers more reliable by use of redundant residues in error detection or correction\cite{kalmykov2022error}.

\begin{algorithm}[b!]
	\renewcommand{\algorithmicrequire}{\textbf{Input:}}
	\caption{: Calculation of $(X * Y) mod (2^n + 1)$}\label{alg:1}
	\begin{algorithmic}
		\REQUIRE Inputs X, Y are in diminished-1 format. $n \geq 1$
		\ENSURE $0\leq{X, Y} < {2^n+1}$
		\STATE 1. Generate partial products: $X\textsubscript{i} Y\textsubscript{j} \, \forall i \in [0, 2], j \in [0, n-1]$
		\STATE 2. Complement ‘i’ most significant partial products and perform a left circular shift for them.
		\STATE 3. Use Quantum 3:2 Compressor to generate the Sum and Carry.
		\STATE 4. For $(i = 3, i\text{++}, i \leq n)$:
		\STATE A. Generate partial products: $X\textsubscript{i} Y\textsubscript{j} \, \forall \, j \in [0, n-1]$
		\STATE B. Complement ‘i’ most significant partial products and perform a left circular shift for them.
		\STATE C. Complement the MSB of Carry of previous stage and perform a left circular shift.
		\STATE D. Add the Sum and Carry generated in (i-1)\textsuperscript{th} iteration along with ‘i’ partial products generated in Step A using Quantum 3:2 Compressor.
		\STATE 5. Complement the MSB of Carry and perform a left circular shift.
		\STATE 6. Add the final Sum and Carry using a carry propagate adder.
		\STATE 7. Complement the resulting Carry and add it back to the sum using a half carry propagate adder to generate $\hspace{7mm} P\textsubscript{i} \; \forall \, i \in [0, n-1]$
		\STATE 8. $P\textsubscript{n} = X\textsubscript{n} \lor Y\textsubscript{n}$
		\STATE 9. If $P\textsubscript{n} = 1, P\textsubscript{i} = 0 \; \forall \, i \in [0, n-1]$
	\end{algorithmic}
	\vspace*{-1mm}
\end{algorithm}

\begin{table*}[!ht]
	\centering
	\caption{Quantum Multiplier Resource Usage Comparison.}
	\label{tab:quantum_multipliers}
	\vspace*{-0.1cm}
	\begin{tabular}{|p{2.5cm}|p{2cm}|p{3.2cm}|p{3.8cm}|c|c|}
		\hline
		\textbf{Quantum Multiplier} & \textbf{Qubit} & \multicolumn{2}{c|}{\textbf{Toffoli}} & \multicolumn{2}{c|}{\textbf{CNOT}} \\
		\cline{3-6}
		& & \textbf{Count} & \textbf{Depth} & \textbf{Count} & \textbf{Depth} \\
		\hline
		\textbf{Proposed Modulo $(2^n + 1)$} & $2n^2 + 4n + 2$ & $2n^2 + 6n - 1$ & $2n^2 + 6n - 2$ & $10n^2 + 16n - 6$ & $10n^2 + 13n - 1$ \\
		\hline
		\textbf{Modulo $(2^n)$ \cite{cho2020quantum}} & $6n - 2 - w(n - 1) - \lfloor\log(n - 1)\rfloor$ & $11n^2 - 22n + 12 - 6(n-1)(w(n - 1) + \lfloor\log(n - 1)\rfloor)$ & $n^2 + 12n - 12 + (n-1)(3\lfloor\log(n - 1)\rfloor + \lfloor\log((n - 1)/3)\rfloor)$ & $4n^2 - 9n + 5$ & $4n - 4$ \\
		\hline
		\textbf{Modulo $(2^n - 1)$ \cite{cho2020quantum}} & $6n - 2$ & $12n^2 - 22n + 11$ & $2n^2 + 7n - 8 + (n-1)(3\lfloor\log(n - 1)\rfloor + \lfloor\log((n - 1)/3)\rfloor)$ & $4n^2 - 4n$ & $4n - 4$ \\
		\hline
		\textbf{Mu{\~n}oz-Coreas et al (Original) \cite{munoz2018quantum}} & $4n + 1$ & $3n^2 - 2$ & $3n^2 - 2$ & $5n^2 - 11n + 6$ & $3n^2 - 5n + 2$ \\
		\hline
		\textbf{Mu{\~n}oz-Coreas et al (QCLA version) \cite{munoz2018quantum}} & $6n - w(n) - \lfloor\log n\rfloor - 1$ & $11n^2 - 15n + 5 - 3(n-1)(w(n) + w(n - 1) + \lfloor\log n\rfloor + \lfloor\log(n - 1)\rfloor)$ & $(n-1)(\lfloor\log n\rfloor + \lfloor\log(n - 1)\rfloor + \lfloor\log n/3\rfloor + \lfloor\log((n-1)/3)\rfloor) + 9n - 8$ & $4n^2 - 9n + 5$ & $4n - 4$ \\
		\hline
	\end{tabular}
	\vspace*{-0.1cm}
	\begin{center}
		\footnotesize
		Log is base 2, and w(n) represents number of 1’s in binary expansion of n.
	\end{center}
	\vspace*{-4mm}
\end{table*}

\section{Proposed Quantum Diminished-1 \\Modulo (2\textsuperscript{n} + 1) Multiplier}
\label{proposed multiplier}
We utilize Algorithm \ref{alg:1} to construct the Quantum Diminished-1 Modulo (2\textsuperscript{n} + 1) Multiplier (QDMM). The two inputs, X = x\textsubscript{n} x\textsubscript{n-1} ... x\textsubscript{0} and Y = y\textsubscript{n} y\textsubscript{n-1} ... y\textsubscript{0}, are (n+1) qubit integers, and their modulo product, P = X * Y = P\textsubscript{n+1} P\textsubscript{n} ... P\textsubscript{0}, is also an (n+1) qubit integer. Algorithm \ref{alg:1} uses a diminished-1 representation to simplify multiplication by a single qubit. To convert a standard integer to diminished-1, we subtract one in the range (1, 2\textsuperscript{n}), with zero represented as 2\textsuperscript{n}. The algorithm optimizes modulo (2\textsuperscript{n} + 1) multiplication using carry-save addition of partial products. It complements and anti-clockwise circular shifts intermediate quantities that exceed the (n-1)\textsuperscript{th} qubit from left to right. Figure \ref{fig:mult_pp} shows the preprocessing and rearrangement of partial products for the (n-1) qubits of inputs X and Y.

These partial products are utilized in Algorithm \ref{alg:1} by passing through the Quantum 3:2 Compressor, which uses one ancilla and provides intermediate Sum and Carry. 
The MSB of each intermediate Carry is inverted and cyclically left-shifted. Ultimately, the final Sum and Carry pair is computed using the Quantum Carry Propagate Adder (QCPA), while the output's Carry is added back into the Sum using half QCPA, yielding the first (n-1) qubits of Product P. To determine if the product is zero, we simply need to calculate the logical OR of the n\textsuperscript{th} qubit of both inputs. In such a case, we employ Toffoli gates to reset the remaining qubits of Product P to $\ket{0}$. Figure \ref{fig:mult_2n+1} illustrates the proposed QDMM for the n=4 configuration, resulting in quantum modulo (2\textsuperscript{4} + 1 = 17) multiplication.

\section{RNS based Distributed Quantum Multiplier}
\label{results}
Table \ref{tab:quantum_multipliers} outlines the quantum resource usage of quantum modulo multipliers used to build Residue Number System (RNS) based distributed quantum multipliers. It breaks down the qubit count, Toffoli and CNOT gate count, and depth of different quantum multiplier designs. 
For quantum modulo (2\textsuperscript{n}-1) and (2\textsuperscript{n}) multipliers, we relied on the designs presented by Cho et al \cite{cho2020quantum}. These quantum modulo multipliers use O(log n) depth Quantum Carry Look Ahead (QCLA) adders, first introduced by Draper et al \cite{draper2004logarithmic}. However, the work by Cho et al. didn't fully account for O(log n) depth, only providing basic complexity based quantities. This required us to recalculate their resource usage estimates.

For comparison with existing quantum multipliers, we selected the T gate count-optimized Quantum Multiplier by Muñoz-Coreas et al. \cite{munoz2018quantum}. This circuit uses a quantum conditional adder that requires very few ancilla and produces no garbage. However, its quantum adder is ripple carry-based, which is inherently less efficient than QCLA-based designs. Therefore, we provide estimates for a QCLA version of the Quantum Multiplier by Muñoz-Coreas et al. As shown in Table \ref{tab:quantum_multipliers}, its Toffoli depth complexity is O(n log n), compared to O(n\textsuperscript{2}) for the original ripple carry version.
\begin{table}[b]
	\vspace*{-0.3cm}
	\centering
	\caption{Quantum Modulo Multiplier Cost Estimates.}
	\vspace*{-0.1cm}
	\label{tab:moduli}
	\begin{adjustbox}{max width=\linewidth}
		\begin{tabular}{|c|c|c|c|c|c|c|c|}
			\hline
			\textbf{n} & \textbf{Type} & \textbf{Mod} & \textbf{Qubit} & \multicolumn{2}{c|}{\textbf{Toffoli}} & \multicolumn{2}{c|}{\textbf{CNOT}} \\
			\cline{5-8}
			& & & & \textbf{Count} & \textbf{Depth} & \textbf{Count} & \textbf{Depth} \\
			\hline
			2 & $2^n-1$ & 3 & 10 & 15 & 12 & 8 & 4 \\
			\hline
			2 & $2^n$ & 4 & 9 & 6 & 14 & 3 & 4 \\
			\hline
			2 & $2^n+1$ & 5 & 18 & 19 & 18 & 66 & 65 \\
			\hline
			3 & $2^n-1$ & 7 & 16 & 53 & 35 & 24 & 8 \\
			\hline
			3 & $2^n$ & 8 & 14 & 21 & 37 & 14 & 8 \\
			\hline
			3 & $2^n+1$ & 9 & 32 & 35 & 34 & 132 & 128 \\
			\hline
			4 & $2^n-1$ & 15 & 22 & 115 & 61 & 48 & 12 \\
			\hline
			4 & $2^n$ & 16 & 19 & 46 & 61 & 33 & 12 \\
			\hline
			4 & $2^n+1$ & 17 & 50 & 55 & 54 & 218 & 211 \\
			\hline
		\end{tabular}
	\end{adjustbox}
\end{table}
\subsection{Cost Analysis}
\vspace*{-0.1cm}
Table \ref{tab:moduli} shows the quantum resource usage for the quantum modulo (2\textsuperscript{n}-1, 2\textsuperscript{n}, 2\textsuperscript{n}+1) multipliers, which are used in this work. The primary observation is that modulo (2\textsuperscript{n}) is the most optimal of the three designs. The proposed modulo (2\textsuperscript{n} + 1) design is better than modulo (2\textsuperscript{n} - 1) in terms of Toffoli count and depth but uses more CNOT gates and qubits.

\begin{table*}[ht!]
	\vspace*{-0.1cm}
	\centering
	\caption{Non-Distributed vs Distributed Quantum Multiplication Resource Consumption Comparison.}
	\vspace*{-0.1cm}
	\label{tab:quantum_multiplication_comparison}
	\begin{tabular}{|c|c|c|c|c|c|c|c|c|c|c|c|c|c|c|}
		\hline
		\multirow{3}{*}{\parbox{1cm}{\centering\textbf{Input\\Size (n)}}} & \multirow{3}{*}{\parbox{1cm}{\centering\textbf{Max.\\Range}}} & \multirow{3}{*}{\parbox{1cm}{\centering\textbf{Output\\Size (2n)}}} & \multicolumn{5}{c|}{\textbf{Non-Distributed Quantum Multiplication}} & \multicolumn{7}{c|}{\textbf{Distributed Quantum Multiplication}} \\
		& & & \multicolumn{5}{c|}{\textbf{(Munoz-Coreas et al (QCLA ver.) }} & \multicolumn{7}{c|}{\textbf{(RNS)}} \\
		\cline{4-8} \cline{9-15}
		& & & \multirow{2}{*}{\textbf{Qubit}} & \multicolumn{2}{c|}{\textbf{Toffoli}} & \multicolumn{2}{c|}{\textbf{CNOT}} & \multirow{2}{*}{\textbf{RNS Set}} & \multirow{2}{*}{\textbf{Range}} & \multirow{2}{*}{\textbf{Qubit}} & \multicolumn{2}{c|}{\textbf{Toffoli}} & \multicolumn{2}{c|}{\textbf{CNOT}} \\
		\cline{5-6} \cline{7-8} \cline{12-13} \cline{14-15}
		& & & & \textbf{Count} & \textbf{Depth} & \textbf{Count} & \textbf{Depth} & & & & \textbf{Count} & \textbf{Depth} & \textbf{Count} & \textbf{Depth} \\
		\hline
		\textbf{3} & 49 & 6 & 14 & 29 & 21 & 14 & 8 & (3, 4, 5) & 59 & 18 & 19 & 18 & 66 & 65 \\
		\hline
		\textbf{4} & 225 & 8 & 20 & 67 & 37 & 33 & 12 & (5, 8, 9) & 359 & 32 & 35 & 37 & 132 & 128 \\
		\hline
		\textbf{5} & 961 & 10 & 25 & 121 & 53 & 60 & 16 & (4, 5, 7, 9) & 1260 & 32 & 53 & 35 & 132 & 128 \\
		\hline
		\textbf{6} & 3969 & 12 & 31 & 191 & 71 & 95 & 20 & (5, 7, 9, 16) & 5040 & 32 & 53 & 61 & 132 & 128 \\
		\hline
		\textbf{7} & 16129 & 14 & 36 & 277 & 91 & 138 & 24 & (7, 9, 16, 17) & 17136 & 50 & 55 & 61 & 218 & 211 \\
		\hline
		\textbf{8} & 65025 & 16 & 43 & 400 & 113 & 189 & 28 & (5, 7, 9, 16, 17) & 85680 & 50 & 55 & 61 & 218 & 211 \\
		\hline
	\end{tabular}
	\vspace*{-0.35cm}
	\label{tab:cost_comparison}
\end{table*}

\begin{table}[hb!]
	\vspace*{-0.3cm}
	\centering
	\caption{Comparative Analysis of Distributed Quantum Multiplication (RNS) with respect to Non-Distributed Quantum Multiplication.}
	\vspace*{-0.1cm}
	\label{tab:perf_improvement}
	\begin{tabular}{|p{0.7cm}|c|c|c|c|c|c|}
		\hline
		\textbf{Output} & \multicolumn{2}{c|}{\textbf{Toffoli Count}} & \multicolumn{2}{c|}{\textbf{Toffoli Depth}} & \multicolumn{2}{c|}{\textbf{T Gate Count}} \\
		\cline{2-7}
		\textbf{Size (2n)} & \textbf{Impr.} & \textbf{Impr. \%} & \textbf{Impr.} & \textbf{Impr. \%} & \textbf{Impr.} & \textbf{Impr. \%} \\
		\hline
		6 & 10 & 34.483 & 3 & 14.286 & 70 & 34.483 \\
		\hline
		8 & 32 & 47.761 & 0 & 0 & 224 & 47.761 \\
		\hline
		10 & 68 & 56.198 & 18 & 33.962 & 476 & 56.198 \\
		\hline
		12 & 138 & 72.251 & 10 & 14.085 & 966 & 72.251 \\
		\hline
		14 & 222 & 80.144 & 30 & 32.967 & 1554 & 80.144 \\
		\hline
		16 & 345 & 86.25 & 52 & 46.018 & 2415 & 86.25 \\
		\hline
	\end{tabular}
	\vspace*{-0.1cm}
\end{table}
This is because it utilizes Quantum 3:2 Compressor as a building block while quantum modulo (2\textsuperscript{n} - 1) multiplier uses QCLA modulo (2\textsuperscript{n} - 1) adder which is garbage less and uses fewer CNOT gates. Table \ref{tab:cost_comparison} shows the equivalent RNS sets for quantum multiplier of size 'n' whose output is of size '2n' and has a maximum range of (2\textsuperscript{n} - 1)*(2\textsuperscript{n} - 1). The non-distributed quantum multiplication is represented by quantum multiplier of Munoz-Coreas et al (QCLA version) \cite{munoz2018quantum}. For RNS based distributed quantum multiplication, we display the maximum quantities among the constituent quantum modulo multipliers across the various quantum resources. This helps in making an objective comparison across the multiple quantum circuits that will be executed in parallel.

Table \ref{tab:perf_improvement} provides the improvement (or decrease) in Toffoli count, depth and T gate count. While Toffoli depth decreases upto 46.018\%, the Toffoli count reduces from 34.483\% to 86.25\%, from 6 qubit to 16 qubit output size respectively. Since T gate usage is seven times in a fault tolerant implementation of Toffoli gate, its decline follows the same trend as that of Toffoli gate count \cite{gosset2013algorithm}.
\vspace{-0.15cm}

\section{Conclusion}
\label{conclusion}
We conclude that the Residue Number System (RNS)-based distributed quantum multiplication offers a promising solution to the scalability challenges of quantum arithmetic operations in Fault Tolerant Quantum (FTQ) computers. This is accomplished by achieving substantial reductions in Toffoli depth, up to 46.018\%, and decreasing T gate counts by 34.483\% to 86.25\% compared to existing non-distributed quantum multipliers for output sizes between 6 and 16 qubits. These improvements become more pronounced as input sizes increase, emphasizing the scalability advantages of the RNS approach. By distributing quantum modulo multiplication circuits across multiple quantum computers or jobs with lower resource requirements, this method addresses the fundamental limitations imposed by high Toffoli depth and T gate usage that have constrained the applicability of quantum multipliers in larger-scale quantum algorithms. In the future, we aim to extend our work to applications and comparisons against more types of quantum multipliers.
\vspace*{-0.1cm}

\bibliographystyle{IEEEtran}
\bibliography{IEEEabrv, references}

\begin{thebibliography}{10}
\providecommand{\url}[1]{#1}
\csname url@samestyle\endcsname
\providecommand{\newblock}{\relax}
\providecommand{\bibinfo}[2]{#2}
\providecommand{\BIBentrySTDinterwordspacing}{\spaceskip=0pt\relax}
\providecommand{\BIBentryALTinterwordstretchfactor}{4}
\providecommand{\BIBentryALTinterwordspacing}{\spaceskip=\fontdimen2\font plus
\BIBentryALTinterwordstretchfactor\fontdimen3\font minus
  \fontdimen4\font\relax}
\providecommand{\BIBforeignlanguage}[2]{{%
\expandafter\ifx\csname l@#1\endcsname\relax
\typeout{** WARNING: IEEEtran.bst: No hyphenation pattern has been}%
\typeout{** loaded for the language `#1'. Using the pattern for}%
\typeout{** the default language instead.}%
\else
\language=\csname l@#1\endcsname
\fi
#2}}
\providecommand{\BIBdecl}{\relax}
\BIBdecl

\bibitem{shor1994algorithms}
P.~W. Shor, ``Algorithms for quantum computation: discrete logarithms and
  factoring,'' in \emph{Proceedings 35th annual symposium on foundations of
  computer science}.\hskip 1em plus 0.5em minus 0.4em\relax Ieee, 1994, pp.
  124--134.

\bibitem{duan2020survey}
B.~Duan, J.~Yuan, C.-H. Yu, J.~Huang, and C.-Y. Hsieh, ``A survey on hhl
  algorithm: From theory to application in quantum machine learning,''
  \emph{Physics Letters A}, vol. 384, no.~24, p. 126595, 2020.

\bibitem{lykov2023fast}
D.~Lykov, R.~Shaydulin, Y.~Sun, Y.~Alexeev, and M.~Pistoia, ``Fast simulation
  of high-depth qaoa circuits,'' in \emph{Proceedings of the SC'23 Workshops of
  The International Conference on High Performance Computing, Network, Storage,
  and Analysis}, 2023, pp. 1443--1451.

\bibitem{munoz2018quantum}
E.~Mu{\~n}oz-Coreas and H.~Thapliyal, ``Quantum circuit design of a t-count
  optimized integer multiplier,'' \emph{IEEE Transactions on Computers},
  vol.~68, no.~5, pp. 729--739, 2018.

\bibitem{mohan2016residue}
P.~A. Mohan, \emph{Residue Number Systems}.\hskip 1em plus 0.5em minus
  0.4em\relax Springer, 2016.

\bibitem{roetteler2017quantum}
M.~Roetteler, M.~Naehrig, K.~M. Svore, and K.~Lauter, ``Quantum resource
  estimates for computing elliptic curve discrete logarithms,'' in
  \emph{Advances in Cryptology--ASIACRYPT 2017}.

\bibitem{putranto2022another}
D.~S.~C. Putranto, R.~W. Wardhani, H.~T. Larasati, and H.~Kim, ``Another
  concrete quantum cryptanalysis of binary elliptic curves,'' \emph{Cryptology
  ePrint Archive}, 2022.

\bibitem{miller2014mapping}
D.~M. Miller, M.~Soeken, and R.~Drechsler, ``Mapping ncv circuits to optimized
  clifford+ t circuits,'' in \emph{Reversible Computation: 6th International
  Conference, RC 2014, Kyoto, Japan, July 10-11, 2014. Proceedings 6}.\hskip
  1em plus 0.5em minus 0.4em\relax Springer, 2014, pp. 163--175.

\bibitem{gosset2013algorithm}
D.~Gosset, V.~Kliuchnikov, M.~Mosca, and V.~Russo, ``An algorithm for the
  t-count,'' \emph{arXiv preprint arXiv:1308.4134}, 2013.

\bibitem{gosset2014algorithm}
------, ``An algorithm for the t-count,'' \emph{Quantum Information \&
  Computation}, vol.~14, no. 15-16, pp. 1261--1276, 2014.

\bibitem{tang2024distributed}
W.~Tang and M.~Martonosi, ``Distributed quantum computing via integrating
  quantum and classical computing,'' \emph{Computer}, vol.~57, no.~4, pp.
  131--136, 2024.

\bibitem{barral2024review}
D.~Barral, F.~J. Cardama, G.~D{\'\i}az, D.~Fa{\'\i}lde, I.~F. Llovo, M.~M.
  Juane, J.~V{\'a}zquez-P{\'e}rez, J.~Villasuso, C.~Pi{\~n}eiro, N.~Costas
  \emph{et~al.}, ``Review of distributed quantum computing. from single qpu to
  high performance quantum computing,'' \emph{arXiv preprint arXiv:2404.01265},
  2024.

\bibitem{gaur2024residue}
B.~Gaur, T.~S. Humble, and H.~Thapliyal, ``Residue number system (rns) based
  distributed quantum addition,'' in \emph{2024 IEEE Computer Society Annual
  Symposium on VLSI (ISVLSI)}, 2024, pp. 595--600.

\bibitem{gaur2024crosstalk}
B.~Gaur and H.~Thapliyal, ``Crosstalk attack resilient rns quantum addition,''
  \emph{arXiv preprint arXiv:2410.23217}, 2024.

\bibitem{draper2004logarithmic}
T.~G. Draper, S.~A. Kutin, E.~M. Rains, and K.~M. Svore, ``A logarithmic-depth
  quantum carry-lookahead adder,'' \emph{arXiv preprint quant-ph/0406142},
  2004.

\bibitem{chevignard2024reducing}
C.~Chevignard, P.-A. Fouque, and A.~Schrottenloher, ``Reducing the number of
  qubits in quantum factoring,'' \emph{Cryptology ePrint Archive}, 2024.

\bibitem{gidney2025factor}
C.~Gidney, ``How to factor 2048 bit rsa integers with less than a million noisy
  qubits,'' \emph{arXiv preprint arXiv:2505.15917}, 2025.

\bibitem{kalmykov2022error}
I.~A. Kalmykov, V.~P. Pashintsev, K.~T. Tyncherov, A.~A. Olenev, and N.~K.
  Chistousov, ``Error-correction coding using polynomial residue number
  system,'' \emph{Applied Sciences}, vol.~12, no.~7, p. 3365, 2022.

\bibitem{cho2020quantum}
S.-M. Cho, A.~Kim, D.~Choi, B.-S. Choi, and S.-H. Seo, ``Quantum modular
  multiplication,'' \emph{Ieee Access}, vol.~8, pp. 213\,244--213\,252, 2020.

\end{thebibliography}

\end{document}